\documentclass[sigconf,natbib=true,nonacm]{acmart}
\usepackage{longtable}
\usepackage{todonotes}
\presetkeys
    {todonotes}
    {inline,backgroundcolor=yellow}{}
\usepackage{multirow}
\usepackage{xspace}
\usepackage{fixltx2e} % For compatibility fixes
\usepackage{stfloats} % For better handling of footnotes in floats

\newcommand{\palpha}{\ensuremath{\text{P}^3\!\alpha}\xspace}
\newcommand{\pbeta}{\ensuremath{\text{RP}^3\!\beta}\xspace}

\newcommand{\easer}{EASE$^R$\xspace}

\usepackage{enumitem}
\setlist[description]{leftmargin=0.2cm,labelindent=0cm}
\usepackage{balance}
\usepackage{hyperref}

\AtBeginDocument{%
  }

\setcopyright{acmlicensed}
\copyrightyear{2025}
\acmYear{2025}

\begin{document}

\title{A Worrying Reproducibility Study of Intent-Aware Recommendation Models}

\author{Faisal Shehzad}
\email{faisal.shehzad@aau.at}
\orcid{1234-5678-9012}
\affiliation{%
  \institution{University of Klagenfurt}
  \city{Klagenfurt}
  \country{Austria}
}

\author{Maurizio Ferrari Dacrema}
\orcid{0000-0001-7103-2788}
\email{maurizio.ferrari@polimi.it}
\affiliation{%
  \institution{Politecnico di Milano}
  \city{Milano}
  \country{Italy}
}

\author{Dietmar Jannach}
\orcid{0000-0002-4698-8507}
\email{dietmar.jannach@aau.at}
\affiliation{%
  \institution{University of Klagenfurt}
  \city{Klagenfurt}
  \country{Austria}
}

\renewcommand{\shortauthors}{Shehzad et al.}

% ========================================
\begin{abstract}
Lately, we have observed a growing interest in \emph{intent-aware} recommender systems (IARS). The promise of such systems is that they are capable of generating better recommendations by predicting and considering the underlying motivations and short-term goals of consumers. From a technical perspective, various sophisticated neural models were recently proposed in this emerging and promising area. In the broader context of complex neural recommendation models, a growing number of research works unfortunately indicates that \emph{(i)} reproducing such works is often difficult and \emph{(ii)} that the true benefits of such models may be limited in reality, e.g., because the reported improvements were obtained through comparisons with untuned or weak baselines. In this work, we investigate if recent research in IARS is similarly affected by such problems. Specifically, we tried to reproduce five contemporary IARS models that were published in top-level outlets, and we benchmarked them against a number of traditional non-neural recommendation models.
In two of the cases, running the provided code with the optimal hyperparameters reported in the paper did not yield the results reported in the paper. Worryingly, we find that all examined IARS approaches are consistently outperformed by at least one traditional model. These findings point to sustained methodological issues and to a pressing need for more rigorous scholarly practices.
\end{abstract}
% ========================================


\maketitle
% ========================================
\section{Introduction}
\label{sec:introduction}
% ========================================
\emph{Intent-aware} recommender systems (IARS) are designed to capture the underlying motivations and goals of users when they use an online service, with the aim of leveraging this information for improved item suggestions~\cite{rafique_developing_2023,jannach2024intent}. In the e-commerce domain, for example, consumers may have different intents when visiting an online shop. A consumer's goal might be to understand the space of options in a certain category, to compare alternatives, to search for ideas, or just hedonic browsing~\cite{shi_improving_2017}. Depending on the current intent, different sets of recommendations may be appropriate. Similarly, in the music domain, users might use a streaming app for different reasons, e.g., to discover new artists, to listen to their favorite tracks, or to find some music to play in a certain situation~\cite{mehrotra_jointly_2019,Kapoor:2015:ILE:2792838.2800172}.  Generally, adapting the recommendations to the assumed user intent has a strong value potential, as evidenced in the field study in the e-commerce domain reported in~\cite{shi_improving_2017}.

Technically, a number of algorithmic approaches to implement intent-awareness in recommender systems have been proposed over the last decade. Early approaches mainly aimed at the diversification of the recommendations, with the goal of covering multiple possible intents through a single recommendation list~\cite{vargas_intent-oriented_2011,wasilewski_intent-aware_2016}. Some other works assume the existence of a fixed set of pre-defined intents in a given domain~\cite{shi_improving_2017,mehrotra_jointly_2019}. Most recently, however, a multitude of \emph{latent intent} modeling approaches were proposed. Such approaches are commonly based on sophisticated deep learning architectures which comprise specific network components designed to model the hidden relationships between observed user interactions and the underlying latent intents. Corresponding proposals have been made both for traditional \emph{top-n} recommendation scenarios, e.g.,~\cite{ma_learning_2019,wang_disentangled_2020,wang_learning_2021}, as well as for sequential recommendation settings, e.g.,~\cite{tanjim_attentive_2020,di_multi-intent_2022,li_multi-intention_2023}, see~\cite{jannach2024intent} for a recent survey on IARS.

Despite the increased sophistication of IARS models and recommendation models in general, and despite the substantial computational complexity and carbon footprint of such models~\cite{Vente2024fromClicks,Spillo2024Towards}, a number of recent studies have indicated that more complex models are not necessarily more effective than longer-existing simpler ones. Ferrari Dacrema et al.~\cite{ferraridacremaetal2019,ferraridacrema2020tois} for example found that in almost all cases they examined, a recent neural recommendation model was outperformed by traditional algorithms based, e.g., on matrix factorization.\footnote{Notably, the examined cases were all published in top-level conferences or journals.} Similar observations of the competitive performance of traditional learning techniques were later reported also in~\cite{Rendle2020NeuMF,RendleKZK22Revisiting,anellitop2022,RevisitingBPR2024}. For the case of sequential recommendation settings, related reports on the effectiveness of simple models---also in practical settings---can be found in ~\cite{latifi2021ins,ludewiglatifiumuai2020,Kouki2020RecSys,Kersbergen2022Serenade,shehzad2024performance}. Overall, while numerous papers claim to advance the state-of-the-art, the achieved progress in this area may actually be very limited. Similar phenomena where ``\emph{improvements don't add up}'' were reported earlier in the field of information retrieval~\cite{Lin:2019:NHC:3308774.3308781,Armstrong:2009:IDA:1645953.1646031} and other application areas of machine learning~\cite{Makridakis2018}.

A number of reasons contribute to the problem. First, the level of reproducibility is generally limited, and still today many papers are published without sharing code and data. Second, different methodological issues and certain bad practices such as using sampled metrics may lead to unreliable results~\cite{Krichene2022Sampled}. Finally, earlier reproducibility studies mentioned above indicate that researchers in many cases do not properly tune the hyperparameters of the baseline models or only consider baselines that are generally weak.

In this paper, we investigate if similar phenomena of ``\emph{phantom progress''}~\cite{ferraridacremaetal2019} can be observed for the emerging and promising area of Intent-Aware Recommender Systems. To that purpose, we have tried to reproduce five recent neural IARS models---all of them published in highly-ranked outlets---and benchmarked them against traditional \emph{top-n} recommendation models. Specifically, as done in~\cite{ferraridacremaetal2019}, we benchmark the considered IARS models using the code, data, and experimental configuration used in the original papers. Our findings are quite worrying, indicating that the mentioned methodological problems in terms of baseline selection and tuning still persist in the community. In all five cases we found that complex IARS models are \emph{not} outperforming existing non-neural techniques if these are properly tuned. Thus, additional measures seem urgently required to improve the scientific rigor in our community.

The paper is organized as follows. Next, in Section~\ref{sec:methodology}, we outline our research methodology. The results of our experiments are then described in Section~\ref{sec:results}. The paper ends with a discussion of implications of our findings.

% ========================================
\section{Methodology}
\label{sec:methodology}
% ========================================
\subsection{Identification of Papers}

In order to avoid cherry-picking individual models for inclusion in our study, we followed a structured approach similar to~\cite{ferraridacremaetal2019}. First, we queried Google Scholar and IEEE Xplore for papers containing the terms `intent' or `intent awareness' along with the keywords `recommend' in the title, abstract, or introduction. We reviewed the retrieved papers for relevance and identified more papers by examining each paper's references. This process left us with 88 articles.\footnotetext{\href{https://recsysevaluation.github.io/IntentAwareRS/}{The list of articles can be found online}. A similar set of papers was identified in a recent survey~\cite{jannach2024intent}.} To further focus our analysis, we only retained works that fulfilled the following criteria: \emph{(i)} the work proposes a model for \emph{top-n} recommendation,
\emph{(ii)} the work was published recently, i.e., during the last four years, \emph{(iii)} it was published in a CORE A* conference or in a journal with an Impact Factor $\geq 5$. Applying these criteria left us with 13 papers.

In the next phase, our goal was to reproduce the results reported in these papers using the artifacts (code and data) that were used in the original papers. We adopt ACM's terminology, where reproducibility means that ``\emph{that an independent group can obtain the same result using the author's own artifacts}.''\footnote{\url{https://www.acm.org/publications/policies/artifact-review-and-badging-current}.} If the authors did not share the artifacts needed for reproducibility, we contacted the authors via email. After waiting for 30 days, we considered
an article to be a candidate for reproduction if it fulfilled the following criteria:

\begin{itemize}
    \item A working version of the code is available or requires minimal modification to make it runnable.
    \item At least one public dataset is used and the authors either share the train-test splits or provide sufficient information in the paper to create the train-test splits from the given dataset.
\end{itemize}

From the 13 papers, we could get the provided code running five with reasonable effort,\footnote{We refer readers to the discussions in~\cite{ferraridacremaetal2019}, where the authors found that shared code repositories are often incomplete, miss relevant parts, or only contain a skeleton of the model.} leading to a  percentage of executable artifacts of 38.46\%. This rate is almost identical to the one observed in a related previous study from 2019~\cite{ferraridacremaetal2019}, indicating that the level of reproducibility may not have increased much since then. Table~\ref{tab:reproduciblity} shows the works providing runnable code and works without code or non-functioning code.

\begin{table}[htbp]
    \caption{Papers with runnable code vs.~papers with missing artifacts or non-runnable code.}
    \label{tab:reproduciblity}
    \centering
    \begin{tabular}{p{3cm}p{3.5cm}}
    Runnable & Missing or non-runnable \\ \hline
    \cite{qian2022intent},\cite{wang2020disentangled},\cite{ren2023disentangled},\cite{zhang2024exploring},\cite{wang2021learning}  & \cite{hu2020graph},\cite{wu2023intent},\cite{li2023topic},\cite{li2023entity},\cite{zhang2023fire},\cite{ma2019learning}, \cite{liu2020octopus},\cite{yang2019intention}
    \end{tabular}
\end{table}

\subsection{Evaluation Methodology}
\paragraph{Benchmarking Approach}
In the literature, two approaches to benchmarking algorithms can be found, each of them having specific advantages. One approach is to integrate all models to be compared in a common evaluation framework, as done, e.g., in~\cite{anellitop2022} or~\cite{ludewiglatifiumuai2020}.
Such an approach ensures that all algorithms are evaluated under identical conditions, e.g., same data splitting, same metrics. Furthermore, such a setting allows us to assess the effectiveness of different models across a collection of datasets beyond the ones that were used for the original paper. The alternative is to evaluate and benchmark existing models using the exact same configuration as was used in the original paper, as done in~\cite{ferraridacremaetal2019}. This approach should allows us to exactly reproduce the results that were reported in the original paper. Since the goal of our present paper is to reproduce the original papers, we adopt this latter approach. Technically, this in particular also means that we rely on the code provided by the authors for the evaluation, i.e., no changes to the evaluation methodology are introduced in our experiments. The downside of this approach is that the reproduced models usually cannot be compared to each other, as there are always small differences in the evaluation methodology across papers. Such a comparison of IARS models is however not the goal of this paper.

Instead, the goal of our work is to understand \emph{(i)} if published results can be reproduced based on the artifacts provided by authors and \emph{(ii)} to what extent recent IARS models are favorable over traditional models. We therefore selected a number of baseline models of different types for our study. The selection is inspired by previous benchmark studies~\cite{ferraridacremaetal2019,anellitop2022}.
The following baselines were considered.

\begin{description}
\item[TopPop:] A non-personalized algorithm that recommends the most popular items to everyone. The popularity of items is calculated based on the number of interactions in the training data.

\item[ItemKNN:]  A traditional collaborative filtering algorithm based on item similarities~\cite{wang2006unifying}. We use cosine similarity in our study. The hyperparameters considered in our study are the neighborhood size, the item weighting strategy, and a shrinking weight for items with fewer interactions~\cite{wang2008probabilistic,bell2007improved}.

\item[UserKNN:]  Like ItemKNN, but considering user similarities instead of item similarities~\cite{Resnick:1994:GOA:192844.192905}.

\item[\palpha:]  A bipartite graph based algorithm that implements a walk between users and items~\cite{cooper2014P3alpha}. Previous work indicates that the approach can be very effective despite its simple nature~\cite{ferraridacremaetal2019}.  The model's hyperparameter are a similarity scaling coefficient $\alpha$ and the neighborhood size.

\item[\pbeta:]  An advanced version of \palpha, using an additional hyperparameter $\beta$ to account for the item popularity.

\item[\easer:]   An ``embarrassingly shallow'' linear model designed for sparse data~\cite{steck2019embarrassingly}. The model has only one hyperparameter and has shown to lead to very competitive performance in previous studies~\cite{anellitop2022,ferraridacrema2020tois}.

\end{description}

We systematically tuned the hyperparameters for all baselines for all experimental configurations. Technically, we used Bayesian optimization and explored 35 to 50 cases, depending on the complexity of the model. The first five cases where selected randomly to reduce the chances of premature convergence. We share all code and data used in our experiment, as well as hyperparameter ranges and best hyperparameters for each dataset online.\footnote{\href {https://recsysevaluation.github.io/IntentAwareRS/}{https://recsysevaluation.github.io/IntentAwareRS/}}

% ========================================
\section{Results}
\label{sec:results}
% ========================================
In this section, we discuss the results for the five IARS models with runnable code ordered by publication date.

\subsection{Disentangled Graph Collaborative Filtering (DGCF)}
\label{sec:DCCF}
\paragraph{Summary} The DGCF model~\cite{wang2020disentangled} was presented at SIGIR '20. The authors highlight that existing models treat user-item interactions as isolated or uniform, neglecting a user's diverse intentions on items, such as shopping for others, scrolling through social media to pass the time, etc. Such uniform approach is assumed to lead to a suboptimal representation of user-item interaction data. The proposed DGCF model consists of two key components to obtain disentanglement: (i) a graph disentangling component, which converts user/item embedding into chunks and attaches an intent to each chunk; (ii) an independence component to enhance the independence between intents.

\paragraph{Datasets and Code}
Experiments were conducted with datasets called Gowalla, Amazon Book and Yelp2018. For all datasets, only users and items with at least ten interactions were retained. For each dataset, an 80/20 rule was adopted for data splitting. For each user, 80\% of the interactions were used for training and the rest was used for testing.\footnote{We note that user-wise splitting may in general lead to data leakage, as future interactions of other users can inform the predictions of the model of a given user.}
The authors shared the train-test splits, which were used to obtain the results reported in the published article.

\paragraph{Original Evaluation}
The authors considered five baselines, including a traditional matrix factorization technique, three graph-based models, and a previous disentanglement  method called MacridVAE. Recall@K and NDCG@K are used as performance measures.  In terms of hyperparameter tuning, the authors mention that for three models they copied the \emph{optimal} hyperparameters from an own previous paper~\cite{Wang2019NGCF}. Looking up the reference, we find that hyperparameters were apparently not fully tuned in~\cite{Wang2019NGCF}, because a central parameter, the embedding size, was not tuned but set to a fixed size. The same fixed size of 64 is also used for two of the graph-based models. Other hyperparameters like the learning rate are tuned. For MacridVAE, embedding sizes were searched between 55 and 75, even though the original paper used an embedding size of 100 (for a different dataset). No code is shared for the baselines, data pre-processing, and hyperparameter tuning.

\paragraph{Our Results}
In Table~\ref{tab:tDGCF} and in subsequent tables we report the results using the metrics and cutoffs used in the original papers. We print numbers in bold when they represent the best results or when a baseline model performs equally or better than the proposed IARS model. \emph{R} stands for Recall, and \emph{N} for NDCG.

\begin{table}[htbp]
\addtolength{\tabcolsep}{-0.1em}
    \caption{Experimental results for DGCF model. DGCF (Paper) refers to the numbers reported in the original paper, DGCF (Code) are the results obtained by running the provided artifacts, DGCF (\cite{Anelli2023Myth}) refers to results obtained in a related evaluation by other researchers.}
    \label{tab:tDGCF}
    \centering
    \begin{tabular}{l l l l l l l}

        \multirow{2}{*}{Methods} & \multicolumn{2}{c}{Gowalla}                              & \multicolumn{2}{c}{Yelp 2018}                                 & \multicolumn{2}{c}{Amazon Book} \\

        ~                       & R@20               & N@20                                 & R@20                  & N@20                            & R@20                    & N@20 \\ \hline
        TopPop                  & 0.013              & 0.031                                & 0.012                 & 0.010                           & 0.002                   &   0.004         \\
        ItemkNN                 & 0.156              & 0.123                                & \textbf{0.064}        & \textbf{0.054}                  & \textbf{0.056}          & \textbf{0.045}            \\
        UserkNN                 & 0.155              & 0.128                                &  0.061                & \textbf{0.050}                  & \textbf{0.051}          & \textbf{0.041}             \\
        \palpha                 & \textbf{0.175}     & 0.146                                & \textbf{0.064}        & \textbf{0.052}                  & \textbf{0.056}          & \textbf{0.043}           \\
        \pbeta                  & \textbf{0.183}     & \textbf{0.152}                       & \textbf{0.067}        & \textbf{0.056}                  & \textbf{0.068}          & \textbf{0.056}           \\
        \easer                  & 0.163              & 0.136                                & \textbf{0.064}        & \textbf{0.052}                  &  \textbf{0.071}         &  \textbf{0.056}                \\ \hline
        DGCF (Paper)            & \textcolor{gray}{0.179}            & \textcolor{gray}{0.152}                              & \textcolor{gray}{0.064}             & \textcolor{gray}{0.052}                           & \textcolor{gray}{0.039}                 & \textcolor{gray}{0.030}                 \\
        DGCF (Code)             & 0.133              & 0.109                                & 0.037                 & 0.028                           & 0.011                   & 0.017                 \\
        DGCF (\cite{Anelli2023Myth}) & 0.173         & 0.147                                & 0.062                  & 0.050                             & 0.038                    & 0.029                 \\

    \end{tabular}
\end{table}
In the table, we report three sets of values for DGCF. \emph{DGCF (Paper)} shows the numbers reported in the original paper. \emph{DGCF (Code)} refers to the numbers when we executed the author-provided code.
Only one set of hyperparameters is provided by the authors, which we used for all datasets. The obtained numbers are much lower than the results reported in the original paper.\footnote{We contacted the authors on this issue without success.}  \emph{DGCF (\cite{Anelli2023Myth})} are performance results obtained by the authors of \cite{Anelli2023Myth}, using a new PyTorch implementation of DGCF  and systematic hyperparameter tuning. We used the best hyperparameters reported in~\cite{Anelli2023Myth} with the original code by the authors, and we then obtained similar results. Comparing the optimal hyperparameters used in~\cite{Anelli2023Myth} and the author-provided ones, we particularly find that using a larger batch size (2000 vs. 1024) may help to improve the performance of the DGCF model. Thus, with proper hyperparameter tuning, the reported results of DGCF can be roughly reproduced, even though the reproduced results are a bit lower.\footnote{We print the numbers reported in the paper in gray and base our comparison on the results that could actually be reproduced.}

However, as also reported in~\cite{Anelli2023Myth}, the simple \pbeta method outperforms DGCF an all metrics and on all datasets. This is true also when comparing the results to the numbers reported in the paper. On the Yelp dataset, all compared methods lead to rather similar results. On the Amazon Book dataset, DGCF is not performing well, and it is actually outperformed by all simple models. We could not run experiments with \easer for the Amazon Book dataset due to hardware constraints. We therefore report the results obtained in~\cite{Zhu2022BARS} instead, which suggest that \easer is performing best on this dataset. The experiments in~\cite{Zhu2022BARS} are based on the same train-test splits as used in the original DGCF paper. The authors of~\cite{Zhu2022BARS} however do not rely on sampled metrics~\cite{Krichene2022Sampled}.

\subsection{Knowledge Graph-based Intent Network (KGIN)}
\label{sec:KGIN}
\paragraph{Summary}
The KGIN approach~\cite{wang2021learning} was presented at TheWebConf '21. KGIN is a recommendation model that uses a knowledge graph which models the user-item interactions from two perspectives: \emph{(i)} identifying the intent behind each user-item interaction at a higher level of granularity and \emph{(ii)} proposing a path-aware aggregation mechanism to integrate information from multiple hops and promote the independence of intents for better prediction and interpretability.

\paragraph{Datasets and Code} Three datasets---Amazon Book, Last-FM and AliBaba-iFashion---were used for performance evaluation. Users and items with less than ten interactions were said to be discarded from the data, but our analysis revealed that users and items with fewer interactions exist in the dataset. An 80/20 data split was applied, and the data splits are shared for all datasets. A leave-one-out splitting strategy was employed to obtain validation data from the shared training data.

\paragraph{Original Evaluation} The authors compare KGIN to six baseline models. One of them is a matrix factorization technique, and the five others are based on Graph Neural Networks and Knowledge Graphs. The embedding size for all models was set to 64 ``\emph{for a fair comparison}''. Again, this is methodologically wrong as the embedding size is a hyperparameter to tune, which can drastically impact model performance~\cite{shehzad2024performance}. Furthermore, for some baselines, fixed values, e.g., batch size and neighborhood size are used. No code is shared for the baselines, data pre-processing, and hyperparameter tuning.

\paragraph{Our Results} The authors provide scripts to train and evaluate the models for all three datasets, stating that these scripts use the best hyperparameters. The results of running the scripts and the performance results for our baselines are shown in Table~\ref{tab:kgnn}.

\begin{table}[htbp]
    \caption{Experimental results for the KGIN model }
    \label{tab:kgnn}
    \centering
    \begin{tabular}{l l l l l }
        \multirow{2}{*}{Methods} & \multicolumn{2}{c}{Amazon Book}                 & \multicolumn{2}{c}{Alibaba-iFashion} \\

        ~                       & R@20             & N@20                        & R@20               & N@20 \\ \hline
        TopPop                  & 0.028            & 0.012                       & 0.031              & 0.016       \\
        ItemkNN                 & \textbf{0.164}   & \textbf{0.097}              & \textbf{0.122}     & \textbf{0.080}        \\
        UserkNN                 & \textbf{0.159}   & \textbf{0.091}              & \textbf{0.101}     & \textbf{0.065}         \\
        \palpha                 & \textbf{0.166}   & \textbf{0.094}              & \textbf{0.069}     & \textbf{0.046}          \\
        \pbeta                  & \textbf{0.144}   & \textbf{0.084}              & \textbf{0.110}     & \textbf{0.073}          \\
        EASE                    & \textbf{0.161}   & \textbf{0.092}              & \textbf{0.120}     & \textbf{0.076}          \\ \hline
        KGIN (Paper)            & \textcolor{gray}{0.168} & \textcolor{gray}{0.091} & \textcolor{gray}{0.114}   & \textcolor{gray}{0.071}                   \\
        KGIN (Code)                   & 0.143              & 0.064                       & 0.102                   & 0.058                \\
         & & & & \\
        \multirow{2}{*}{}       & \multicolumn{2}{c}{Last-FM (leaking)}                       & \multicolumn{2}{c}{Last-FM*} \\

        TopPop                  & 0.022                   & 0.019                    & 0.015                   & 0.009                \\
        ItemkNN                 & 0.182                   & \textbf{0.180}           & \textbf{0.234}          & \textbf{0.202}        \\
        UserkNN                 & 0.130                   & 0.143                    & \textbf{0.200}          & \textbf{0.168}         \\
        $P_{3}\alpha$           & 0.194                   & \textbf{0.196}           & \textbf{0.239}          & \textbf{0.213}          \\
        $RP_{3}\beta$           & 0.189                   & \textbf{0.187}           &  \textbf{0.253}         & \textbf{0.233}                    \\
        EASE                    & 0.185                   & \textbf{0.184}           & \textbf{0.227}          & \textbf{0.191}     \\ \hline
        KGIN (Paper)            & \textcolor{gray}{0.097} & \textcolor{gray}{0.0848}           & --                   & --                   \\
        KGIN (Code)             & \textbf{0.202}          & 0.173           & 0.163                   & 0.092                   \\
    \end{tabular}
\end{table}

With \emph{KGIN (Paper)}, we again refer to the results in the paper, and \emph{KGIN (Code)} are the numbers obtained by running the scripts by the authors which are said to lead to the best performance. The obtained numbers are often substantially lower, with the exception for Recall for the Last-FM dataset, which is more than twice as high as in the paper. We contacted the authors on this issue without success, leading to a situation where the reported results are not reproducible. Given the surprising results for the Last-FM dataset, we investigated the provided train-test splits to find a significant data leakage issue, affecting 17,290 out of 23,566 users. We therefore created the union of the provided train-test splits, i.e., removed the duplicates, and created our own 80/20 split, leading to what we denote the Last-FM* dataset. The results for the different models on this dataset are also shown in Table~\ref{tab:kgnn}.\footnote{We used the provided hyperparameters for Last-FM for the Last-FM* dataset.}

Comparing the performance of our baselines and KGIN, we find that many of our baselines outperform KGIN on all three (correct) datasets. In particular the UserKNN and ItemKNN work very well on all datasets.  We note that for all datasets and measurements except for Recall@20 for the Amazon Book dataset, at least one baseline model leads to better results than the non-reproducible higher ones reported in the paper.

\subsection{Intent Disentanglement and Feature Self-supervision (IDS4NR)}
\label{sec:IDS4NR}

\paragraph{Summary} The IDS4NR model~\cite{qian2022intent} was published in IEEE TKDE in 2022. The authors present a mechanism that optimizes accuracy and novelty by disentangling the users' intent for popular items (conformity bias) and their personal interests. To model the relationships between users and items different traditional recommendation models can be used as a \emph{backbone}. Three backbones are considered in the paper, matrix factorization (named LFM), neural collaborative filtering (NCF), and collaborative metric learning (CML).

\paragraph{Datasets and Code} The authors use the MovieLens-100k dataset as well as the Amazon Music and Amazon Beauty datasets for their experiments. The datasets are augmented with side information. Only items and users with at least five interactions are retained. Training and test splits are provided. No code for data preprocessing, baselines, and tuning is shared.

\paragraph{Original Evaluation} The evaluation considers Recall as an accuracy metric, as well as Coverage and Novelty as beyond-accuracy metric. An F-score metric is computed as the harmonic mean of the three metrics. One set of hyperparameters for the proposed model is shared. The optimization goal is not reported. The embedding dimension for all models is set to 50, which, as mentioned, is a major methodological problem. Hyperparameters for the baselines are said to be taken from the original papers or fine-tuned---without further specification---when they were unknown.

\paragraph{Our Results}
We could run the code for two of three versions of the proposed model (when using LFM and NCF as backbones) using the shared artifacts and the provided hyperparameters. Running the code for the CML backbone required the specification of a number of parameters that were not shared in the paper or in the provided code.
Inquiries to the authors were without success.

\begin{table*}[htbp]
    \caption{Experimental results for IDS4NR model}
    \label{tab:idsnr}
    \centering
    \begin{tabular}{l l l l l l l l l}
        \multirow{2}{*}{Methods} & \multicolumn{8}{c}{MovieLens} \\
        ~                        & R@5            & R@10           & Cov@5            & Cov@10          & Nov@5          & Nov@10            & F1@5              & F1@10   \\ \hline
       TopPop                    & 0.052          & 0.087          & 0.025	           & 0.040	         & 0.023	      & 0.046	          & 0.049	          & 0.092 \\
        ItemKNN                  & \textbf{0.090} & \textbf{0.141} & 0.057            & 0.080           & 0.024          & 0.048             & 0.085             & 0.139  \\
        UserKNN                  & 0.073          & 0.110          & 0.036            & 0.055           & 0.023          & 0.047             & 0.061             & 0.110 \\
        $P_{3}\alpha$            & \textbf{0.105} & \textbf{0.162} & 0.104            & 0.153           & 0.024          & 0.049             & 0.136             & 0.224 \\
        $RP_{3}\beta$            & \textbf{0.106} & \textbf{0.174} & 0.160            & 0.220           & 0.025          & 0.050             & 0.197             & 0.303 \\
        EASE                     & \textbf{0.119} & \textbf{0.185} & 0.179            & 0.250           & 0.025          & 0.050             & 0.217             & 0.338 \\ \hline

        IDNSR\_LFM   (Code)      & 0.080          & 0.138           & \textbf{0.367}  & \textbf{0.486}  & \textbf{0.030} & \textbf{0.062}    & \textbf{0.431}    & \textbf{0.637} \\
        IDNSR\_LFM  (Paper)      & \textcolor{gray}{0.087}          & \textcolor{gray}{0.143}     & \textcolor{gray}{\textbf{0.441}}   & \textcolor{gray}{\textbf{ 0.572}}  & \textcolor{gray}{0.3870} & \textcolor{gray}{0.390}          & \textcolor{gray}{0.049}        & \textcolor{gray}{0.087} \\
        IDSNR\_NCF   (Code)      & 0.084          & 0.138           & \textbf{0.390}  & \textbf{0.539 } & \textbf{0.030} & \textbf{0.060}    & \textbf{0.454}     & \textbf{0.696} \\
        IDSNR\_NCF  (Paper)      & \textcolor{gray}{0.085}          & \textcolor{gray}{0.138}        & \textcolor{gray}{0.467}         & \textcolor{gray}{0.615}           & \textcolor{gray}{0.390}  & \textcolor{gray}{ 0.385}          & \textcolor{gray}{0.049}            & \textcolor{gray}{0.086} \\

         %\hline
        \multirow{2}{*}{} & \multicolumn{8}{c}{Amazon Music} \\ \hline
        TopPop                  & 0.020          & 0.033          & 0.004          & 0.007          & 0.011          & 0.023          & 0.015          & 0.031 \\
        ItemKNN                 & \textbf{0.153} & \textbf{0.211} & \textbf{0.831} & 0.918          & 0.016          & 0.031          & \textbf{0.885} & 1.035 \\
        UserkNN                 & \textbf{0.146} & \textbf{0.201} & 0.468          & 0.638          & 0.014          & 0.028          & 0.501          & 0.721 \\
        $P_{3}\alpha$           & \textbf{0.155} & \textbf{0.212} & 0.571          & 0.724          & 0.014          & 0.029          & 0.610          & 0.816 \\
        $RP_{3}\beta$           & \textbf{0.160} & \textbf{0.219} & 0.782          & 0.897          & 0.015          & 0.030          & 0.832          & 1.007 \\
        EASE                    & \textbf{0.158} & \textbf{0.214} & 0.553          & 0.721          & 0.014          & 0.029          & 0.591          & 0.812 \\ \hline

        IDNSR\_LFM  (Code)      & 0.061          & 0.106          & 0.392          & 0.524          & 0.014          & 0.029          & 0.423          & 0.599 \\
        IDNSR\_LFM (Paper)      &\textcolor{gray}{0.089}             & \textcolor{gray}{0.136}            & \textcolor{gray}{0.685}         & \textcolor{gray}{ 0.8324}           & \textcolor{gray}{0.500} & \textcolor{gray}{\textbf{0.498}}        & \textcolor{gray}{0.071}            & \textcolor{gray}{0.115} \\
        IDSNR\_NCF  (Code)      & 0.067          & 0.112          & 0.800          & \textbf{0.985} & \textbf{0.017} & \textbf{0.033} & 0.856          & \textbf{1.117} \\
        IDSNR\_NCF (Paper)      & \textcolor{gray}{0.090}            & \textcolor{gray}{0.137}           & \textcolor{gray}{0.903}   & \textcolor{gray}{0.974}  & \textcolor{gray}{0.535} & \textcolor{gray}{0.524} & \textcolor{gray}{0.085}             & \textcolor{gray}{0.128} \\

        \multirow{2}{*}{} & \multicolumn{8}{c}{Amazon Beauty} \\ \hline
        TopPop                  & 0.006          & 0.010          & 0.002         & 0.003         & 0.007           & 0.015         & 0.009         & 0.018 \\
        ItemkNN                 & \textbf{0.041} & \textbf{0.058} & 0.350         & 0.512         & 0.008           & 0.017         & 0.368         & 0.555 \\
        UserkNN                 & \textbf{0.042} & \textbf{0.062} & 0.447         & 0.628         & 0.009           & 0.018         & 0.467         & 0.679 \\
        $P_{3}\alpha$           & \textbf{0.043} & \textbf{0.062} & 0.481         & 0.658         & 0.009           & 0.018         & 0.502         & 0.711 \\
        $RP_{3}\beta$           & \textbf{0.045} & \textbf{0.064} & \textbf{0.656}& \textbf{0.785}& 0.009           & 0.018         & \textbf{0.683}& \textbf{0.847} \\
        EASE                    & \textbf{0.050} & \textbf{0.067} & 0.456         & 0.639         & 0.009           & 0.018         & 0.477         & 0.691 \\ \hline

        IDNSR\_LFM  (Code)      & 0.019          & 0.032          & 0.255          & 0.403        & \textbf{0.011}  & 0.022         & 0.274         & 0.453 \\
        IDNSR\_LFM (Paper)      & \textcolor{gray}{0.027}         & \textcolor{gray}{0.042}                     & \textcolor{gray}{0.602}       & \textcolor{gray}{0.711}           & \textcolor{gray}{0.400} & \textcolor{gray}{0.412} & \textcolor{gray}{0.019}             & \textcolor{gray}{0.032}          \\

        IDSNR\_NCF  (Code)      & 0.023          & 0.037          & 0.581          & 0.770        & 0.010           & \textbf{0.021} & 0.609        & 0.840 \\
        IDSNR\_NCF (Paper)      & \textcolor{gray}{0.025}            & \textcolor{gray}{0.041}            & \textcolor{gray}{0.902}            & \textcolor{gray}{0.973}           & \textcolor{gray}{0.538} & \textcolor{gray}{0.536} & \textcolor{gray}{0.025}             & \textcolor{gray}{0.041}  \\ \hline
    \end{tabular}
\end{table*}

The results of executing the provided code and the results for the baselines are shown in Table~\ref{tab:idsnr}. We observe that the numbers reported in the paper largely deviate from the numbers we obtain when running the provided code. In terms of accuracy measures, the obtained numbers are about 70\% higher than what is reported in the paper for the MovieLens dataset. For the other two datasets, the Recall values are in contrast substantially lower. Thus, with the provided artifacts, we could not reproduce the results. In any case, independent of which performance results for IDSNR we consider, i.e., from the paper or ours, all our traditional and simpler baselines lead to higher Recall values and outperform IDSNR. The same holds for the F-measure that combines Recall with Coverage and Novelty.

For Coverage and Novelty, we also observe that our results often deviate largely from the numbers reported in the paper. IDSNR is generally very strong in terms of these metrics, compared to the baselines. However, considering the lower F-measure value, it becomes clear that increasing Coverage and Novelty comes at a too high price in terms of accuracy.

\subsection{Disentangled Contrastive Collaborative Filtering (DCCF)}
\label{sec:DCCF}
\paragraph{Summary} The DCCF model~\cite{ren2023disentangled} was presented at SIGIR '23. The model implements intent disentanglement combined with adaptive self-supervised augmentation. The latter component shall help the model to be more robust against noise.

\paragraph{Datasets and Code} The authors evaluated their model on the Gowalla, Amazon Book, and Tmall datasets. Differently from other related works like the DGCF model discussed above, the datasets were apparently not preprocessed, or preprocessed differently than in previous works. This can make a comparison with previous works difficult.\footnote{Often, authors also do not share which \emph{version} of a dataset is used. There are, for example, now three versions of the Amazon Book dataset available by now at~\url{https://cseweb.ucsd.edu/~jmcauley/datasets/amazon_v2/}.} The authors however share the training and test data that were used in the experiments. No information about the splitting procedure are provided. No code for preprocessing, baselines, or hyperparameter tuning is shared.

\paragraph{Original Evaluation} The authors benchmark their method against as many as 12 baseline models from different families. Hyperparameter ranges for the proposed model are described. For the baseline models, a fixed embedding size of 32 and batch size of 1024 are used, again representing a major methodological issue. No information about the selection of these values is provided. Additional hyperparameter values for the considered baselines and the target metric for tuning are not reported either. The authors use Recall and NDCG as evaluation measures. The choice of the cutoff length of 40 is unusual and not explained in the paper.

\paragraph{Our Results} We could run the provided code, leading to the results shown in Table~\ref{tab:dccf}. We observe that the obtained numbers are very close to those reported in the paper, sometimes a bit higher, sometimes a bit lower. We therefore consider this work reproducible. We however find that all of the considered baseline models outperform the proposed DCCF model on all datasets.\footnote{Again, we could not run the \easer model on the largest Amazon Book dataset due to hardware constraints.} Among our baselines, no unique winner can be identified, and the relative performance of different algorithms depends on dataset characteristics.

\begin{table}[htbp]
    \caption{Experimental results for the DCCF model}
    \label{tab:dccf}
    \centering
    \begin{tabular}{l l l l l }
        \multirow{2}{*}{Methods} & \multicolumn{4}{c}{Gowalla} \\
        ~                        & R@20               & R@40                    & N@20               & N@40       \\ \hline
        TopPop               & 0.025                & 0.037                     & 0.013                 & 0.017         \\
        UserKNN                  & \textbf{0.213}       & \textbf{0.295}            & \textbf{0.132}        & \textbf{0.153} \\
        ItemKNN                  & \textbf{0.235}       & \textbf{0.321}            & \textbf{0.146}        & \textbf{0.168} \\
        $P_{3}\alpha$            & \textbf{0.231}       & \textbf{0.317}            & \textbf{0.141}        & \textbf{0.164}   \\
        $RP_{3}\beta$            &  \textbf{0.245}      &  \textbf{0.337}           & \textbf{0.149}        & \textbf{0.173}     \\
        EASE                     &  \textbf{0.218}      &  \textbf{0.303}           & \textbf{0.132}        & \textbf{0.154}      \\
        DCCF  (Code)                   & 0.189	            & 0.266	                    &   0.112	            &  0.132               \\
        DCCF (Paper)             & \textcolor{gray}{0.187}	            & \textcolor{gray}{0.264}	                    &   \textcolor{gray}{0.112}	            & \textcolor{gray}{0.132}              \vspace{5pt} \\
        \multirow{2}{*}{} & \multicolumn{4}{c}{Amazon Book} \\ \hline
        TopPop              & 0.009                &   0.016                   & 0.007                 & 0.009             \\
        UserKNN                 & \textbf{0.126}       & \textbf{0.172}            & \textbf{0.104}        & \textbf{0.119}            \\
        ItemKNN                 & \textbf{0.152}       & \textbf{0.205}            & \textbf{0.127}        &  \textbf{0.144}           \\
        $P_{3}\alpha$           & \textbf{0.133}       & \textbf{0.188}            & \textbf{0.107}        & \textbf{0.124}            \\
        $RP_{3}\beta$           & \textbf{0.141}       & \textbf{0.190}            & \textbf{0.117}        &  \textbf{0.133}           \\
        EASE                    &  -                   &    -                     &  -                  &     -        \\
        DCCF  (Code)                  & 0.092	               &   0.138	               & 0.071	               &   0.086                                \\
        DCCF (Paper)                    & \textcolor{gray}{0.088}	               &   \textcolor{gray}{0.134} 	               & \textcolor{gray}{0.068}	               &   \textcolor{gray}{0.082}   \vspace{5pt}                              \\
        \multirow{2}{*}{} & \multicolumn{4}{c}{Tmall} \\ \hline
        TopPop              & 0.010                & 0.017                     & 0.007                 & 0.009            \\
        UserKNN                 & \textbf{0.087}       & \textbf{0.131}            & \textbf{0.063}        & \textbf{0.079}            \\
        ItemKNN                 & \textbf{0.110}       & \textbf{0.163}            & \textbf{0.080}        & \textbf{0.099}            \\
        $P_{3}\alpha$           & \textbf{0.109}       & \textbf{0.162}            & \textbf{0.080}        & \textbf{0.099}            \\
        $RP_{3}\beta$           & \textbf{0.116}       & \textbf{0.173}            & \textbf{0.084}        & \textbf{0.104}           \\
        EASE                    &  \textbf{0.093}      & \textbf{0.143}            & \textbf{0.067}        & \textbf{0.084}            \\
        DCCF  (Code)                  &   0.064	           & 0.101	                   &  0.045	               & 0.058             \\
        DCCF (Paper)            &   \textcolor{gray}{0.066}             & \textcolor{gray}{0.104}	                   &  \textcolor{gray}{0.046}	               & \textcolor{gray}{0.059}             \\ \hline
    \end{tabular}
\end{table}

\subsection{Bilateral Intent-guided Graph Collaborative Filtering (BIGCF)}
\label{sec:BIGCF}
\paragraph{Summary} The BIGCF approach~\cite{zhang2024exploring} was presented at SIGIR '24. It aims at separating a user's individual intent, i.e., personal preferences, from collective intent, which the authors describe as overall awareness. Conceptually, the work is thus related to the IDS4NR approach discussed above. Technically, the model is based on Graph Convolution Networks and Graph Contrastive Learning.

\paragraph{Datasets and Code} Datasets from Gowalla, Amazon Book, and Tmall datasets are used for evaluation. The same version of the datasets is used as for the DCCF model discussed above. The train-test splits used for the experiments are shared by the authors. No further information is provided how the splits were exactly created and which parts of the data were used to tune the hyperparameters. The target metric for tuning is not reported either. No code for preprocessing, baselines, or hyperparameter tuning is shared.

\paragraph{Original Evaluation} The authors closely follow the evaluation setup that was used in~\cite{ren2023disentangled} to evaluate the DCCF approach. Specifically, they also rely on the problematic approach to fix central hyperparameters like the embedding size to a (small) fixed value for all compared models. For the hyperparameters of the proposed model, the search ranges for the hyperparameters are reported. While the evaluation involves three datasets, only one set of optimal values is highlighted in the paper. In the shared code repository, scripts are provided to run the model on three datasets. For these scripts, specific values for the number of epochs and the regularization factor are provided. It remains unclear how the number of epochs was chosen, as no use of a validation set is mentioned.\footnote{The study in~\cite{Sun2020Arewe} indicates that a substantial fraction of the literature (37\% of the examined papers) were tuned on test data.}

For the compared methods, the authors mention that they used hyperparameters from the original papers or used grid search. No further details are provided.  The proposed BIGCF model is then benchmarked against fifteen baseline models from the literature. There is a significant overlap in terms of the baselines with respect to the evaluation of DCCF. Some baselines from the DCCF paper are however not considered, while others are added. For the baselines that are overlapping, the reported numbers are in most cases exactly the same as those reported in~\cite{ren2023disentangled}. An exception is the NGCF model, where the results in the papers differ.\footnote{The authors of the BIGFC rely on the evaluation code used for the DCCF model. Similar results are thus generally expected.}

\begin{table}[ht!]
    \caption{Experimental results for the BIGCF model}
    \label{tab:bigcf}
    \centering
    \begin{tabular}{l l l l l }
    %\hline
        \multirow{2}{*}{Methods} & \multicolumn{4}{c}{Gowalla} \\
        ~                        & R@20                 & R@40                      & N@20                  & N@40       \\ \hline
        TopPop               & 0.025                & 0.037                     & 0.013                 & 0.017          \\
        UserKNN                  & \textbf{0.213}       & \textbf{0.295}            & \textbf{0.132}        & \textbf{0.153} \\
        ItemKNN                  & \textbf{0.235}       & \textbf{0.321}            & \textbf{0.146}        & \textbf{0.168} \\
        $P_{3}\alpha$            & \textbf{0.231}       & \textbf{0.317}            & \textbf{0.141}        & \textbf{0.164}   \\
        $RP_{3}\beta$            &  \textbf{0.245}      &  \textbf{0.337}           & \textbf{0.149}        & \textbf{0.173}     \\
        EASE                     &  \textbf{0.218}      &  \textbf{0.303}           & \textbf{0.132}        & \textbf{0.154}    \\ \hline
        BIGCF  (Code)                  & 0.206	            & 0.289	                    &   0.123	            &  0.145           \\
        BIGCF (Paper)            & \textcolor{gray}{0.208}               & \textcolor{gray}{0.288}                    & \textcolor{gray}{0.124} & \textcolor{gray}{0.145}   \\ %\hline

        \multirow{2}{*}{} & \multicolumn{4}{c}{Amazon Book} \\ \hline
        TopPop              & 0.009                &   0.016                   & 0.007                 & 0.009                    \\
        UserKNN                 & \textbf{0.126}       & \textbf{0.172}            & \textbf{0.104}        & \textbf{0.119}            \\
        ItemKNN                 & \textbf{0.152}       & \textbf{0.205}            & \textbf{0.127}        &  \textbf{0.144}           \\
        $P_{3}\alpha$           & \textbf{0.133}       & \textbf{0.188}            & \textbf{0.107}        & \textbf{0.124}            \\
        $RP_{3}\beta$           & \textbf{0.141}       & \textbf{0.190}            & \textbf{0.117}        &  \textbf{0.133}           \\
        EASE                    &  -                   &    -                      &  -                    &     -        \\ \hline
        BIGCF  (Code)                 & 0.099	               & 0.147	                   & 0.076	               &   0.091          \\
        BIGCF (Paper)  & \textcolor{gray}{0.098} & \textcolor{gray}{0.146} & \textcolor{gray}{0.076} & \textcolor{gray}{0.091}         \\ %\hline

        \multirow{2}{*}{} & \multicolumn{4}{c}{Tmall} \\ \hline
        TopPop              & 0.010                & 0.017                     & 0.007                 & 0.009            \\
        UserKNN                 & \textbf{0.087}       & \textbf{0.131}            & \textbf{0.063}        & \textbf{0.079}     \\
        ItemKNN                 & \textbf{0.110}       & \textbf{0.163}            & \textbf{0.080}        & \textbf{0.099}       \\
        $P_{3}\alpha$           & \textbf{0.109}       & \textbf{0.162}            & \textbf{0.080}        & \textbf{0.099}            \\
        $RP_{3}\beta$           & \textbf{0.116}       & \textbf{0.173}            & \textbf{0.084}        & \textbf{0.104}           \\
        EASE                    &  \textbf{0.093}      & \textbf{0.143}            & \textbf{0.067}        & \textbf{0.084}            \\ \hline
        BIGCF  (Code)                 & 0.075	               & 0.117	                   &  0.053	               & 0.068             \\
        BIGCF (Paper)           & \textcolor{gray}{0.075}               & \textcolor{gray}{0.116}                    & \textcolor{gray}{0.053}                & \textcolor{gray}{0.068}\\

    \end{tabular}
\end{table}

\paragraph{Our Results} Running the scripts provided by the authors, we obtained the numbers reported in Table~\ref{tab:bigcf}. We observe that the numbers reported in the paper are very close to those obtained when running the code provided to the authors, meaning that the work is reproducible. However, as for the DCCF model, which was evaluated using the same experimental configuration, we find that all simple baselines outperform the BIGCF model on all datasets and metrics. The \pbeta algorithm is consistently among the best-performing models.

Regarding the scripts provided by the authors, we note that the number of epochs is provided as a hyperparameter for model training for each dataset individually. It is however unclear how the optimal number of epochs was determined. In the provided code, we observe that the performance of the model on the test set is observed during training for each epoch, but this information is not used for testing. We re-ran the experiments with an \emph{early-stopping} strategy. These additional experiments showed that early-stopping often leads to substantially lower accuracy values for the Gowalla and Amazon Book datasets. A proper choice of the number of epochs seems therefore essential for the performance of this model. The number of epochs should however be determined based on a validation dataset.

% ========================================
\section{Discussion}
\label{sec:discussion}
In this section, we first summarize our observations regarding reproducibility of research published in this area. Then, we will discuss methodological issues that we encountered in our reproducibility study, and who these issues may prevent us from making progress.

\paragraph{Reproducibility}
Questions of reproducibility in recommender systems and in information retrieval have been reported again and again in the literature and have been around for at least fifteen years \cite{Armstrong:2009:IDA:1645953.1646031,Ekstrand2011Rethinking,Konstan2013Toward,saidComparativeRecommenderSystem2014}. More than ten years ago, related questions were discussed in a dedicated workshop at ACM RecSys.
About five years ago, a reproducibility study involving deep learning models~\cite{ferraridacremaetal2019} revealed that \emph{(i)} the general level of reproducibility of published research is low and \emph{(ii)} that then-recent deep learning models were actually not as effective in offline studies as one might assume.

The observations made in our present study in the emerging and highly-promising recent area of intent-aware recommender systems are sobering. At least in the studied area of IARS, we found that only five of the thirteen identified relevant papers shared artifacts that are needed for reproducibility. Moreover, as our study showed, only for a subset of them it was then possible to actually reproduce the results that were reported in the papers based on the provided artifacts and documentation.
\begin{raggedright}
Overall, reproducing research in recommender systems therefore remains to be difficult. Besides the issue of lacking reproducibility artifacts and documentation of the proposed models, we found that none of the examined papers shared the codes that were used to obtain the results for the baselines. Codes for hyperparameter tuning and data preprocessing---all required to ensure full reproducibility---were entirely missing as well. Moreover, we found that authors are often not very responsive when approached for guidance in case of reproducibility issues. All in all, this situation is more than worrying, in particular when taking into account that our study focused on papers that were published at high-level venues for which we may assume the highest standards of scientific rigor.
\end{raggedright}
Clearly, the general level of reproducibility in recommender systems research has increased in the last decade. Sharing code was very uncommon fifteen years ago. Nowadays, more and more publication venues encourage authors to share their code and data, and questions related to reproducibility are becoming common in review forms of conferences and journals. However, it seems that the measures that were taken so far are insufficient, and more awareness is needed among various stakeholders to improve the situation~\cite{bauer2023overcoming}. Ultimately, ensuring reproducibility should in most cases be technically feasible, e.g., through containerization or virtualization. Furthermore, various reproducibility guidelines are around for at least ten years~\cite{Konstan2013Toward}. Currently, however, there still seem to be insufficient incentives for researchers to prepare working reproducibility packages.

\paragraph{Methodological Issues and Progress}
The existence of fundamental methodological issues in information retrieval and recommender systems are not new either, considering for example Armstrong et al.'s findings from fifteen years ago~\cite{Armstrong:2009:IDA:1645953.1646031}. Various types of methodological flaws and bad practices were identified over the years, and our study shows that these problems seem to persist in recent research on IARS. The most important issues include the comparison of new models against baselines that were not properly tuned and setups where the potential effectiveness and expressive power of the baselines is artificially restricted by limiting the size of the user and item embeddings. Furthermore, the choice of the baselines also matters. In the examined works, most baselines are rather recent neural models, and only in a few cases more traditional models, e.g., based on matrix factorization, are included. This choice may lead to a continued \emph{propagation of weak baselines}~\cite{ferraridacremaetal2019}.

Furthermore, the establishment of what represents the \emph{state-of-the-art} in terms of a small set of (recent) models that perform well across many use cases and datasets remains difficult and maybe impossible. The relative performance of algorithms may strongly depend on the characteristics of the dataset, the specifics of the evaluation protocol, and how hyperparameter tuning was done~\cite{shehzad2023everyone}. In our study, we found that a range of datasets was used---sometimes even using the same name but applying different preprocessing steps---and that there is also no consistency in terms of cut-off lengths. Specifics of the data splitting approach, which may affect performance outcomes, are also often missing.

Considering the specific papers that were investigated in our study, we may fear that the progress that is made even through the latest models may in fact be quite limited. Despite their conceptual simplicity, we found that longer-existing approaches are highly competitive with the latest complex models, despite the often very high computational costs of such models.

\paragraph{Computational Complexity Analysis}
To illustrate the aspect of computational complexity, we consider the running times for the KGIN model discussed above for the largest dataset, Alibaba iFashion, that was used in the original paper for evaluation. The dataset comprises about 114,000 users, 30,000 items, and about 1.7M interactions. We recall that the Netflix Prize dataset that was released fifteen years ago contained 100M interactions. The running times for the KGIN model on our hardware\footnote{Hardware specifications: NVIDIA RTX A4000, GPU RAM: 16 GB, and CPU RAM: 125 GB} are shown in Table~\ref{tab:performance_metrics-KGIN}. We report the time needed to train the model with the given set of hyperparameters (T-Time), the time needed to make predictions for all users (P-Time), and the time needed to make a prediction for a single user.

In Table~\ref{tab:performance_metrics-KGIN}, we observe that the time needed to train the KGIN model is orders of magnitude higher than our considered baselines. We recall that for this dataset, the decade-old itemKNN approach worked best in terms of accuracy. Running the preprocessing steps\footnote{There is no traditional `training' for nearest-neighbor model. Therefore, we report the time that is needed to setup and fill internal data structures for similarity computations.} for ItemKNN takes less than three minutes, while training KGIN on the GPU may take days to weeks on our hardware. At prediction time, the KGIN model is also 50\% slower than ItemKNN. As discussed in Section~\ref{sec:KGIN}, the performance results that we obtained by running the code provided for the KGIN model were lower than reported in the paper.\footnote{The ItemKNN method is however at least on par with the KGIN model when considering the results in the original paper.} Still, even if the performance of ItemKNN would be slightly surpassed by KGIN or another complex model, the question arises if the immense computational effort and corresponding carbon footprint would be justified for a modest gain in offline accuracy. In particular, this question has to be addressed bearing in mind that the offline-online gap in recommender systems has been discussed for at least a decade~\cite{Levy2013PainGain}, where it is more than unclear if gains in offline accuracy would ever translate into more successful systems in practice.

Having said that, we in no way claim that research in complex models should be dismissed. In contrast, there are various reports of applying such models successfully in practice for several years, e.g.,~\cite{Covingon2016Youtube,DBLP:journals/aim/SteckBELRB21}. Furthermore, there are other reproducibility studies that find that complex models can actually outperform traditional baselines, even if these baselines are extensively tuned, see, e.g., the study by Anelli et al.~\cite{Anelli2023Myth} on models based on Graph Collaborative Filtering. What our study focuses on instead are continuing issues of academic research practices, which seem to continue to prevent us from making faster progress.

\begin{table}[h!]
\centering
\caption{Comparison of training (T-Time) and prediction times (P-Time) for different models for the Alibaba-iFashion dataset.}
\begin{tabular}{l|c|c|c}
Model      & T-Time (s)    & P-Time (s)     & Avg. P-Time/User (ms) \\ \hline
TopPop          & 0.022         & 172.280        & 1.501 \\
ItemkNN         & 140.771       & 201.694        & 1.757 \\
UserkNN         & 381.969       & 162.805        & 1.418 \\
$P_{3}\alpha$   & 37.596        & 170.240        & 1.483 \\
$RP_{3}\beta$   & 38.748        & 173.301        & 1.510 \\
EASE            & 316.519       & 206.727        & 1.801 \\
KGIN            & 117,360.426   & 362.042        & 3.15 \\
\end{tabular}
\label{tab:performance_metrics-KGIN}
\end{table}

\paragraph{Limitations}
Our research is focused on the recent area of intent-aware recommender systems and to a limited set of five works that were published with artifacts for reproducibility and which we identified through a systematic process. While this process allows us to gauge the level of reproducibility in this particular area, more research is needed in the future to assess if similar methodological issues exist in other works in IARS. This includes approaches that rely on explicitly specified domain-specific intents and works that focus on sequential recommendation problems, see~\cite{jannach2024intent}. Furthermore, our study does not allow us to make conclusions about algorithms research in recommender systems in general, or about specific families of models, e.g., graph-based ones. In terms of the baselines considered in our study, we so far only included \easer as a non-neural model-based technique. The exploration of traditional models based on matrix factorization, like ALS or BPR, which both were found to be competitive with recent models~\cite{RendleKZK22Revisiting,RevisitingBPR2024}, is part of our ongoing and future work.

Like comparable previous reproducibility studies, we did \emph{not} further tune the hyperparameter of the proposed IARS model. Instead, our assumption is that the hyperparameters that were shared by the authors correspond to the optimal ones that they found while developing and tuning the models. Given the computational complexity of some of these models, the authors of the original papers may have however been limited in the number of hyperparameter configurations they could explore. Thus, it may be easily possible that the performance of the proposed IARS models can be further improved. We however recall here that such an additional tuning process is not the focus of our current study and would require substantial computational resources. Instead, the goals of our study were \emph{(i)} to assess if the results reported in the original papers can be reproduced using only the artifacts (including documentation) shared by the authors\footnote{A different approach was apparently taken in~\cite{Anelli2023Myth}, who partially relied on alternative implementations of the models, e.g., in the case of the DGCF model. Reimplementing a model may however come with its own risks~\cite{saidComparativeRecommenderSystem2014,HidasiC23ThirdParty}.}, and \emph{(ii)} how these models compare to well-tuned traditional models.

% ========================================
\section{Conclusion \& Outlook}
\label{sec:conclusion}
We have studied questions of reproducibility and progress in the emerging area of intent-aware recommender systems. Worryingly, our analyses point to continuing methodological issues in this area. Our findings thus call for further action to increase awareness in the community about these issues, intensified education of all involved stakeholders, and better incentives for academic scholars so that we are able to achieve research results that are more reliable and better reproducible in the future. The problems discussed in our work are not new~\cite{Konstan2013Toward}, and we for example observe certain improvements over the last decade in particular in terms of sharing of code and data. However, we believe that we as a community still have to strive harder to achieve more rigor in our scientific practices in the future.

\end{document}